\documentclass[intlimits,a4paper,11pt]{article}

\usepackage[eqsecnum,english]{kioj4}

\journalsectionnoimage{}

\issue{2021}{3}{cam}

\title[New features in MathPartner 2021]{NEW FEATURES IN MATHPARTNER 2021}

\authorI[Malaschonok G.I.]{Malaschonok Gennadi \thanks{1111}}
\authorIpos{D.Sc.}
 \authorIemailEnvelope{malaschonok@gmail.com} 
\authorII[Seliverstov A.V.]{Seliverstov Alexandr}
\authorIIpos{Ph.D.}
\authorIIemail{slvstv@iitp.ru}

\begin{document}

    \maketitle

    \begin{abstract}
We introduce new features in the MathPartner service that have recently become available to users. We highlight the functions for calculating both arithmetic-geometric mean and geometric-harmonic mean. They allow calculating complete elliptic integrals of the first kind. They are useful for solving many physics problems, for example, one can calculate the period of a simple pendulum. Next, one can calculate the modified arithmetic-geometric mean proposed by Semjon Adlaj. Consequently, one can calculate the complete elliptic integrals of the second kind as well as the circumference of an ellipse. Furthermore, one can also calculate the Sylvester matrices of the first and the second kind. Thus, by means of a few strings, one can calculate the resultant of two polynomials as well as the discriminant of a binary form. Some new matrix functions are also added. So,  today the list of matrix functions includes the transpose, adjugate, conjugate, inverse, generalized inverse, and pseudo inverse of a matrix, the matrix determinant, the kernel, the echelon form, the characteristic polynomial, the Bruhat decomposition, the triangular LSU decomposition, which is an exact block recursive LU decomposition, the QR block recursive decomposition, and the singular value decomposition. In addition, two block-recursive functions have been implemented for calculating the Cholesky decomposition of symmetric positive-definite matrices: one function for sparse matrices with the standard multiplication algorithm and another function for dense matrices with multiplication according to the Winograd--Strassen algorithm. The linear programming problems can be solved too. So, the MathPartner service has become better and handy. It is freely available at http://mathpar.ukma.edu.ua/ as well as at http://mathpar.com/ .
        \autocitationexample
    \end{abstract}

\section{INTRODUCTION}

The MathPartner service is useful at school, university, and work~\cite{M17a,M17b}. It can help you to solve problems in mathematical analysis, algebra, geometry, physics, and more. You can operate with functions and functional matrices, to obtain the exact numerical and analytical solutions and solutions in which the numerical coefficients have a required accuracy. Today it is available at \url{http://mathpar.ukma.edu.ua/} as well as at \url{http://mathpar.com/}.

We present some new features and improvements. In particular, you can calculate the arithmetic-geometric mean and its modification to calculate the complete elliptic integrals of the first and the second kind~\cite{A12,A19}. Thus, you can calculate the circumference of an ellipse as well as the period of a pendulum. Another application is also proposed to compute packing properties~\cite{LL90,MMMB00,XGW13}. One can also calculate the Sylvester matrix~\cite{A89,AMV16} as well as the resultant of two polynomials. Some new matrix functions have been implemented too~\cite{G11,M13,M19,M20}.

\section{SIX MEANS AND THE COMPLETE ELLIPTIC INTEGRALS} 

Given two non-negative numbers $x$ and $y$, one can define their arithmetic, geometric and harmonic means as $\frac{x+y}{2}$, $\sqrt{xy}$, and $\frac{2xy}{x+y}$, respectively. 
Moreover, $\mathbf{AGM}{(x,y)}$ denotes the arithmetic-geometric mean of $x$ and $y$. It was defined by Johann Carl Friedrich Gauss at the end of the 18th century. $\mathbf{GHM}{(x,y)}$ denotes the geometric-harmonic mean of $x$ and $y$. At last, $\mathbf{MAGM}{(x,y)}$ denotes the modified arithmetic-geometric mean of $x$ and $y$. It is defined by Semjon Adlaj~\cite{A12,A19}. Every mean is a symmetric homogeneous function in their two variables $x$ and $y$. 
In contrast to well-known means, $\mathbf{AGM}{(x,y)}$, $\mathbf{GHM}{(x,y)}$, and $\mathbf{MAGM}{(x,y)}$ are calculated iteratively. 

The arithmetic-geometric mean $\mathbf{AGM}{(x,y)}$ is equal to the limit of both sequences $x_n$ and $y_n$, where $x_0=x$, $y_0=y$, $x_{n+1}=\frac{1}{2}(x_n+y_n)$, and $y_{n+1}=\sqrt{x_ny_n}$.

In the same way, the geometric-harmonic mean $\mathbf{GHM}{(x,y)}$ is equal to the limit of both sequences $x_n$ and $y_n$, where $x_0=x$, $y_0=y$, $x_{n+1}=\sqrt{x_ny_n}$, and $y_{n+1}=\frac{2x_ny_n}{x_n+y_n}$. Note that $\mathbf{AGM}{(x,y)}\mathbf{GHM}{(x,y)}=xy$.

The modified arithmetic-geometric mean $\mathbf{MAGM}{(x,y)}$ is equal to the limit of the sequence $x_n$, where $x_0=x$, $y_0=y$, $z_0=0$, $x_{n+1}=\frac{x_n+y_n}{2}$, $y_{n+1}=z_n+\sqrt{(x_n-z_n)(y_n-z_n)}$, and $z_{n+1}=z_n-\sqrt{(x_n-z_n)(y_n-z_n)}$.

For example, let us run the commands, where functions begin with the symbol $\backslash$, SPACE denotes the ring of coefficients, and FLOATPOS denotes the number of decimal places

\centerline{\tt SPACE=R64[]; FLOATPOS=3; a=$\backslash$AGM(1,5); g=$\backslash$GHM(1,5); m=$\backslash$MAGM(1,5); [a,g,m];}
The output is equal to $[2.604, 1.920, 2.611]$.

These means are applicable, in particular, to calculate the complete elliptic integrals of the first and second kind. Let us use the parameter $0\le k\le 1$. 

The complete elliptic integral of the first kind $K(k)$ is defined as
$$K(k)=\int_0^1\frac{dt}{\sqrt{(1-t^2)(1-k^2t^2)}}$$
It can be computed in terms of the arithmetic-geometric mean:
$$K(k)=\frac{\pi}{2\mathbf{AGM}(1,\sqrt{1-k^2})}$$
On the other hand, for $k<1$, it can be computed in terms of the geometric-harmonic mean:
$$K(k)=\frac{\pi}{2}\mathbf{GHM}\left(1,\frac{1}{\sqrt{1-k^2}}\right)$$

The complete elliptic integral of the second kind $E(k)$ is defined as
$$E(k)=\int_0^1\sqrt{\frac{1-k^2t^2}{1-t^2}}dt$$
It can be computed in terms of the modified arithmetic-geometric mean:
$$E(k)=K(k)\mathbf{MAGM}(1,1-k^2)$$

The circumference of an ellipse is equal to
$$2\pi\frac{\mathbf{MAGM}(a^2,b^2)}{\mathbf{AGM}(a,b)},$$
where the semi-major and semi-minor axes are denoted $a$ and $b$. 

On the other hand, $\pi$ can be expressed as 
$$\pi=\frac{\left(\mathbf{AGM}(1,\sqrt{2})\right)^2}{\mathbf{MAGM}(1,2)-1}$$
So, to calculate $\pi$ one can run the commands

\centerline{\tt SPACE = R[]; FLOATPOS = 24; w = $\backslash$sqrt(2); ($\backslash$AGM(1,w)){\textasciicircum}2/($\backslash$MAGM(1,2)-1);}
Every argument of functions $\mathbf{AGM}()$, $\mathbf{MAGM}()$, and $\mathbf{GHM}()$ must be either a number or a variable, i.e., compound expressions cannot be used as arguments. 

Let a point mass be suspended from a pivot with a massless cord. The length of the pendulum is denoted by $L$. It swings under gravitational acceleration $g = 9.80665 m/s^2$. The maximum angle that the pendulum swings away from the vertical, called the amplitude, is denoted by $\theta_0$. 
One can find the period $T$ of the pendulum using the arithmetic-geometric mean
$$T=\frac{2\pi}{\mathbf{AGM}(1,\cos(\theta_0/2))}\sqrt{\frac{L}{g}}$$
If $L=1 m$ and $\theta_0=120^{\circ}$, then $T=2.7546 s$.  To calculate the period one can run the commands

\centerline{\tt SPACE = R64[]; FLOATPOS = 4; L = 1; g = 9.80665;}
\centerline{\tt T = $\backslash$value(2*$\backslash$pi*$\backslash$sqrt\{L/g\}/($\backslash$AGM(1, 0.5));}
On the other hand, if $\theta_0$ is small, then the period is equal to $2.0064 s$.

Note that we use the $\mathbf{value}()$ command to evaluate $\pi$. 

\section{THE SYLVESTER MATRICES, THE RESULTANT, AND THE DISCRIMINANT}

Let us consider two univariate polynomials $f(x)$ and $g(x)$, where $\deg(f)=n$, $\deg(g)=m$, and $m\leq n$ hold. James Joseph Sylvester introduced two matrices associated to $f(x)$ and $g(x)$. Please, refer to~\cite{A89,AMV16}. More precisely, there are two different Sylvester matrices associated with two univariate polynomials. Let us denote $f(x)=f_nx^n+\cdots+f_1x+f_0$ and $g(x)=g_mx^m+\cdots+g+1x+g_0$. The Sylvester matrix of the first kind was introduced in 1840 \cite{S40}. It is the $(n+m)\times(n+m)$ matrix. Its determinant is called the resultant of $f$ and $g$. For example, if $f=x^3+px+q$ and $g=3x^2+p$, then the Sylvester matrix of the first kind is equal to
$$\left(\begin{array}{ccccc}
1 & 0 & p & q & 0\\
0 & 1 & 0 & p & q\\
3 & 0 & p & 0 & 0\\
0 & 3 & 0 & p & 0\\
0 & 0 & 3 & 0 & p\\
\end{array}\right)$$
and its determinant equals $4p^3+27q^2$, i.e., it is the opposite of the discriminant of $f$.

The Sylvester matrix of the second kind was introduced in 1853 as an improvement of the Sturm theory~\cite{S53}. It is the $(2n)\times(2n)$ matrix, where $n\geq m$. The first and the second rows are
$$\left(\begin{array}{ccccccccc}
f_n & \cdots & f_{m+1} & f_m & \cdots & f_0 & 0 & \cdots & 0\\
0 & \cdots & 0 & g_m & \cdots & g_0& 0 & \cdots & 0\\
\end{array}\right)$$
The next pair is the first pair, shifted one column to the right; the first elements in the two rows are zero. The remaining rows are obtained the same way as above.
For example, if $f=x^3+px+q$ and $g=3x^2+p$, then the Sylvester matrix of the second kind is equal to
$$\left(\begin{array}{cccccc}
1 & 0 & p & q & 0 & 0\\
0 & 3 & 0 & p & 0 & 0\\
0 & 1 & 0 & p & q & 0\\
0 & 0 & 3 & 0 & p & 0\\
0 & 0 & 1 & 0 & p & q\\
0 & 0 & 0 & 3 & 0 & p\\
\end{array}\right)$$
Of course, if the resultant vanishes, then the determinant of the Sylvester matrix of the second kind vanishes too.

The Sylvester matrix of the first kind can be calculated in MathPartner by a ternary function called $\mathbf{sylvester}(\cdot,\cdot,0)$, where the third argument is equal to zero. In the same way, the Sylvester matrix of the second kind can be calculated in MathPartner by a ternary function called $\mathbf{sylvester}(\cdot,\cdot,1)$, where the third argument is not equal to zero. The first and the second arguments are univariate polynomials, for example, $f(x)$ and $g(x)$. The variable must be the last one in the list of variables. For example, if the polynomials over the ring of integers depend on parameters $p$ and $q$, then the declaration in MathPartner can be {\tt SPACE =  Z[p,q,x]}.

The resultant of two univariate polynomials can be calculated as $\mathbf{resultant}(f,g)$. The variable must be the last one in the list of variables. For example, let us run

\centerline{\tt SPACE = Z[a, b, c, x]; f = a*x{\textasciicircum}2+b*x+c; g = 2*a*x+b; $\backslash$resultant(f, g);}
The output is equal to $4ca^2-b^2a$.

The discriminant of a univariate polynomial $f(x)=f_dx^d+\cdots+f_0$ is equal to
$$\mathbf{discriminant}(f)=\frac{(-1)^{d(d-1)/2}}{f_d}\mathbf{resultant}(f,f^{\prime})$$
The discriminant can be calculated immediately. For example,  

\centerline{\tt SPACE = Z[a, b, c, x]; f = a*x{\textasciicircum}2+b*x+c; $\backslash$discriminant(f);}
The output is equal to $-4ca+b^2$. There exists another way to calculate the discriminant of the univariate polynomial $x^2+bx+c$, where $b$ and $c$ are parameters.

\centerline{\tt SPACE = Z[b, c, x]; f = x{\textasciicircum}2+b*x+c; -$\backslash$det($\backslash$sylvester(f, $\backslash$D(f, x), 0));}
The output is equal to $-4c+b^2$. Of course, $\mathbf{D}(f, x)$ calculates the first derivative of $f$.

\section{SYSTEMS OF ALGEBRAIC EQUATIONS}

Let us show an application of the resultant of two univariate polynomials. For this purpose, we consider a system of two polynomial equations in two variables and eliminate a variable. Of course, variable elimination can be done by computing a Gröbner basis. So, there exists another way to solve a system of algebraic equations. Unfortunately, the Gröbner basis approach is sometimes very complicated. Contrariwise, the approach based on the resultant is often more effective. Let us consider the system
$$\left\{
\begin{array}{rcl}
x^2+y^2&=&1\\
2x^2+xy+y^2&=&1\\
\end{array}\right.$$
In this case, solutions to the system correspond to intersection points of the circle and the ellipse.
Let us consider two univariate polynomials depending on one parameter~$x$
$$\begin{array}{rcl}
f(y)&=&x^2+y^2-1\\
g(y)&=&2x^2+xy+y^2-1\\
\end{array}$$
Its resultant is equal to $2y^4-3y^2+1$. On the other hand, the Gröbner basis for the reverse lexicographical ordering consists of two polynomials $x-2y^3+2y$ and $2y^4-3y^2+1$. The second polynomial is equal to the resultant. Thus, every solution to the system satisfies the equation $2y^4-3y^2+1=0$. So, one can eliminate this variable. There exist four solutions to the equation $y_1=-1$, $y_2=-\sqrt{2}/2$, $y_3=\sqrt{2}/2$, and $y_4=1$. The corresponding values of $x$ are $x_1=0$, $x_1=\sqrt{2}/2$, $x_1=-\sqrt{2}/2$, and $x_4=0$. So, there are four intersection points. 

Next, let us show the corresponding program in MathPartner. 
The Gröbner basis of a polynomial ideal can be obtained due to Bruno Buchberger. The algorithm is implemented as $\mathbf{groebnerB}()$. The same basis can be calculated using a matrix algorithm that is  similar to the F4 algorithm. It is implemented as $\mathbf{groebner}()$. The ordering is reverse lexicographical. Note that functions should begin with the symbol $\backslash$. 

\centerline{\tt SPACE = Z[y, x]; f = x{\textasciicircum}2+y{\textasciicircum}2 -1; g = 2*x{\textasciicircum}2+x*y+y{\textasciicircum}2 -1; $\backslash$groebner(f, g);}
The output consists of two polynomials $[x-2y^3+2y, 2y^4-3y^2+1]$.
Another way to calculate the resultant is to run the commands

\centerline{\tt SPACE = Z[y, x]; f = x{\textasciicircum}2+y{\textasciicircum}2 -1; g = 2*x{\textasciicircum}2+x*y+y{\textasciicircum}2 -1; $\backslash$det($\backslash$sylvester(f, g, 0));}
The output consists of one univariate polynomial $2y^4-3y^2+1$. Of course, one can run

\centerline{\tt SPACE = Z[y, x]; f = x{\textasciicircum}2+y{\textasciicircum}2 -1; g = 2*x{\textasciicircum}2+x*y+y{\textasciicircum}2 -1; $\backslash$resultant(f, g);}

To calculate roots of a polynomial one can run $\mathbf{solve}()$. 

\centerline{\tt SPACE = Q[y]; $\backslash$solve(2*y{\textasciicircum}4 -3*y{\textasciicircum}2+1 = 0);}
The output consists of four numbers expressed in radicals $[-1, \sqrt{2}/2, ((-1)\cdot\sqrt{2}/2), 1]$.

Let us run the same command over the field of real numbers. We recommend using option {\tt SPACE = R64[y]}. It denotes the set of $64$-bit floating-point numbers with $52$-digit mantissa, $11$-bit exponent, and one sign bit.

\centerline{\tt SPACE = R64[y]; $\backslash$solve(2*y{\textasciicircum}4 -3*y{\textasciicircum}2+1 = 0);}
The output consists of four floating-point numbers $[1.00, -1.00, 0.71, -0.71]$.

Of course, systems of linear algebraic equations can be solved with $\mathbf{solve}()$. For example,

\centerline{\tt SPACE = Q[]; M = [[1, 2], [3, 1]]; b = [5, 5]; $\backslash$solve(M, b);} 
The output is equal to $[1, 2]^T $. There exists another way 

\centerline{\tt SPACE = Q[x, y]; $\backslash$solve([x+2*y = 5, 3*x+y = 5]);} 
The output is equal to $[1, 2]$.

Moreover, one can solve a system of inequalities in one variable. For example,

\centerline{\tt SPACE = Q[x]; $\backslash$solve([x{\textasciicircum}2+4*x -5 > 0, x{\textasciicircum}2 -2*x -8 < 0]);}
The output is equal to $(1, 4)$. In the next  example

\centerline{\tt SPACE = Q[x]; $\backslash$solve([x < 0, x > 2]);}
The output is equal to the empty set $\emptyset$.

\section{THE GREATEST COMMON DIVISOR OF TWO POLYNOMIALS}

In this section we shall consider polynomials over either the field of rational numbers or the ring of integers. 
The problem of calculating the greatest common divisor of two polynomials is important for symbolic computations, in particular, over a finite extension of the field of rational numbers~\cite{AS20,D18,HL21}. Unfortunately, the bit complexity of the Euclidean algorithm is exponential. There exists a polynomial upper bound on the number of arithmetic operations. But the size of a product of integers at intermediate steps can be very large. For some discussion about the computational complexity of powers of integers refer to~\cite{KS20}. 

A modified algorithm based on subresultant residues had been proposed by J.J.~Sylvester~\cite{S53} and later improved by Walter Habicht~\cite{H48} and   Alkiviadis Akritas \cite{A89}.  

The main result was obtained by Brown. He found a way to compute the subresultant PRS without using matrix reduction. He proposed to modify the Euclidean algorithm, reducing all coefficients by common factors so that they coincide with the subresultant PRS \cite{B78}. This algorithm is applied in MathPartner to compute the GCD of two polynomials. This approach was further developed in work \cite{AMV16}.

 To calculate the greatest common divisor one can run $\mathbf{GCD}(f,g)$; for example,

\centerline{\tt SPACE = Z[x]; $\backslash$GCD(9*x, 6*x+6);}
The output is equal to $3$. To calculate Bézout coefficients one can run $\mathbf{extendedGCD}(f,g)$. The least common multiple can be calculated too.

\centerline{\tt SPACE = Z[x]; $\backslash$LCM(9*x, 6*x+6);}
The output is equal to $18x^2+18x$.

\section{MATRIX FUNCTIONS}

Today the list of matrix functions includes the transpose, adjugate, conjugate, inverse, generalized inverse, and pseudo inverse of a matrix, the matrix determinant, the kernel, the matrix echelon form, the characteristic polynomial, the Bruhat decomposition, the triangular LSU decomposition, which is an exact block recursive LU decomposition, the QR block recursive decomposition, and the singular value decomposition. In addition, two block-recursive functions have been implemented for calculating the Cholesky decomposition of symmetric positive definite matrices: one function for sparse matrices with the standard multiplication algorithm and another function for dense matrices with multiplication according to the Winograd--Strassen algorithm. The linear programming problems can be solved too.

For a given matrix $A$, the pseudo inverse of $A$ is a matrix $A^{-}$ satisfying both equalities $AA^{-}A=A$ and $A^{-}AA^{-}=A^{-}$. Furthermore, the generalized inverse Moore--Penrose  $A^{+}$ satisfies four equalities $AA^{+}A=A$, $A^{+}AA^{+}=A^{+}$, $(A^{+}A)^T=A^{+}A$, and $(AA^{+})^T=AA^{+}$.
If $A$ is a square non-degenerate matrix, then three inverses of $A$ coincide, i.e., $A^{-1}=A^{-}=A^{+}$. If a $n\times m$ matrix $A$ can be decomposed as $A=BC$, where $B$ is a $n\times k$ matrix, $C$ is a $k\times m$ matrix, and $\mathrm{rank}(A) =\mathrm{rank}(B) =\mathrm{rank}(C) = k$, then 
$A^{+}= C^T(CC^T)^{-1}(B^TB)^{-1}B^T$. This idea was expressed by Vera Nikolaevna Kublanovskaya~\cite{K66}. About big matrices refer to~\cite{MT21}.

For a given matrix $A$, one can calculate:
\begin{itemize}
\item 
The transpose: $\mathbf{transpose}(A)$ or $A^{T}$;
\item
The conjugate: $\mathbf{conjugate}(A)$ or $A^{\ast}$;
\item
The matrix echelon form: $\mathbf{toEchelonForm}(A)$;
\item
The kernel: $\mathbf{kernel}(A)$;
\item 
The rank: $\mathbf{rank}(A)$; 
\item
The determinant: $\mathbf{det}(A)$; 
\item
The inverse: $\mathbf{inverse}(A)$ or $A^{-1}$;
\item
The adjugate: $\mathbf{adjoint}(A)$ or $A^{\star}$;
\item
The generalized inverse Moore--Penrose: $\mathbf{genInverse}(A)$ or $A^{+}$; 
\item
The pseudo inverse: $\mathbf{pseudoInverse}(A)$; 
\item
The closure: $\mathbf{closure}(A)$ or $A^{\times}$. The closure of a matrix $A$ is equal to the sum of matrices $I+A+A^2+A^3+\cdots$. For the classical algebras it is equivalent to $(I-A)^{-1}$.
\end{itemize}
To calculate the characteristic polynomial of a matrix $A$, you should work over the ring of polynomials in some new variable and run $\mathbf{charPolynom}(A)$. For example, let us run the commands

\centerline{\tt SPACE=Z[x]; M=[[1, 2], [3, 5]]; f=$\backslash$charPolynom(M);}
The output is equal to $f = x^2-6x-1$.

Let us take a closer look at some types of decomposition.

\subsection{The Bruhat decomposition}

To calculate the Bruhat decomposition of a matrix $A$ one can run $\mathbf{BruhatDecomposition}(A)$. 
The result consists of three matrices $[V,D,U]$, where both $V$ and $U$ are upper-triangular matrices, $D$ is a permutation matrix multiplied by the inverse of the diagonal matrix~\cite{M13}. If all entries of the matrix $A$ are elements of commutative domain $R$, then all entries of matrices $V$, $D^{-1}$, and $U$ belong to the same domain $R$. Let us consider a $2\times 2$ matrix over $\mathbb{Z}$. For example,
$$M=\left(\begin{array}{cc}
-1&2\\
1&1\\
\end{array}\right)$$
Let us run the commands

\centerline{\tt SPACE = Z[]; M = [[-1, 2], [1, 1]]; $\backslash$BruhatDecomposition(M);}
The output consists of three matrices
$$\left[\left(\begin{array}{cc}
3 & -1\\
0 & 1\\
\end{array}\right),
\left(\begin{array}{cc}
0 & 1/3\\
1 & 0\\
\end{array}\right),
\left(\begin{array}{cc}
1 & 1\\
0 & 3\\
\end{array}\right)\right]$$
An entry of the middle matrix $D$ is not integer, but the inverse matrix has integer entries.
$$D^{-1}=\left(\begin{array}{cc}
0 & 1\\
3 & 0\\
\end{array}\right)$$

\subsection{The LSU decomposition}

The LSU decomposition of a matrix $A$ can be calculated by means of the command $\mathbf{LSU}(A)$. 
The result consists of three matrices $[L,S,U]$, where $L$ is a lower-triangular matrix, $U$ is an upper-triangular matrix, $S$ is a permutation matrix multiplied by the inverse of a diagonal matrix. If all entries  of the matrix $A$ belong to a commutative domain $R$, then all entries of matrices $L$, $S^{-1}$, and $U$ belong to the same domain $R$, refer to~\cite{M20}.  Let us consider an example, where $M$ is a $2\times 2$ matrix.
$$M=\left(\begin{array}{cc}
1&2\\
3&1\\
\end{array}\right)$$
Let us run the commands

\centerline{\tt SPACE = Z[]; M = [[1, 2], [3, 1]]; $\backslash$LSU(M);}
The output consists of three matrices
$$\left[\left(\begin{array}{cc}
1 & 0\\
3 & -5\\
\end{array}\right),
\left(\begin{array}{cc}
1 & 0\\
0 & -1/5\\
\end{array}\right),
\left(\begin{array}{cc}
1 & 2\\
0 & -5\\
\end{array}\right)\right]$$
Both first and third matrices are triangular matrices over $\mathbb{Z}$. The middle matrix $S$ has a rational entry, but the inverse matrix is defined over $\mathbb{Z}$. 
$$S^{-1}=
\left(\begin{array}{cc}
1 & 0\\
0 & -5\\
\end{array}\right)$$
To calculate the LSU decomposition of $A$ together with decomposition of the pseudo inverse  $A^{-}=WSM$, one can run the command $\mathbf{LSUWMdet}(A)$. The result consists of five matrices and determinant of the largest non-degenerate corner block $[L,S,U,W,M,\det]$, where $L$ and $U$ are  lower and upper triangular matrices, $S$ is a truncated weighted permutation matrix, $SM$ and $WS$ are  lower and upper triangular matrices. Moreover, $A=LSU$ and $A^{-}=WSM$. If entries of the matrix $A$ belong to a commutative domain, then all matrices, except for $S$, also belong to this domain. Let us run the commands

\centerline{\tt SPACE = Z[]; M = [[1, 2], [3, 1]]; $\backslash$LSUWMdet(M);}
The output consists of 
$$\left[\left(\begin{array}{cc}
1 & 0\\
3 & -5\\
\end{array}\right),
\left(\begin{array}{cc}
1 & 0\\
0 & -1/5\\
\end{array}\right),
\left(\begin{array}{cc}
1 & 2\\
0 & -5\\
\end{array}\right),
\left(\begin{array}{cc}
-5 & 10\\
0 & -5\\
\end{array}\right),
\left(\begin{array}{cc}
-5 & 0\\
15 & -5\\
\end{array}\right),
[[-5]]
\right]$$
Of course, three of these matrices coincide with three matrices in the previous example.

Next, let us consider a matrix over the ring $\mathbb{Z}[x,y]$
$$M=\left(\begin{array}{cc}
y & x\\
x & y\\
\end{array}\right)$$ 
and run the commands

\centerline{\tt SPACE = Z[x, y]; M = [[y, x], [x, y]]; $\backslash$LSU(M);}
The output consists of three matrices
$$\left[\left(\begin{array}{cc}
y & 0\\
x & y^2-x^2\\
\end{array}\right),
\left(\begin{array}{cc}
1/y & 0\\
0 & 1/\left(y^3-yx^2\right)\\
\end{array}\right),
\left(\begin{array}{cc}
y & x\\
0 & y^2-x^2\\
\end{array}\right)\right]$$
Entries of the middle matrix $S$ are rational functions. All entries of the matrices $L$, $S^{-1}$, and $U$ are polynomials over $\mathbb{Z}$.

\subsection{The QR block recursive decomposition}

Let us consider a $2^{k} \times 2^{k}$ matrix $A$ over the field of reals. The QR decomposition of $A$ can be calculated by means of the command $\mathbf{QR}(A)$. Note that if  the order is not equal to $2^k$ for any integer $k$, then the algorithm does not work because it is based on block recursion~\cite{M19}. Let us consider an example, where $M$ is a $2\times 2$ matrix.
$$M=\left(\begin{array}{cc}
1&2\\
3&1\\
\end{array}\right)$$
Let us run the commands

\centerline{\tt SPACE = R64[]; M = [[1, 2], [3, 1]]; $\backslash$QR(M);}
The output consists of two matrices
$$\left[\left(\begin{array}{cc}
0.32 & -0.95\\
0.95 & 0.32\\
\end{array}\right),
\left(\begin{array}{cc}
3.16 & 1.58\\
0 & -1.58\\
\end{array}\right)\right]$$
The first matrix is orthogonal. The second matrix is upper-triangular. Their product is equal to the initial matrix $M$.

\subsection{The singular value decomposition}

To calculate the singular value decomposition (SVD) of a matrix $A$, one can run $\mathbf{SVD}(A)$. As a result, three matrices $[U,D,V]$ will be calculated. The matrices $U$ and $V$ are unitary, the matrix $D$ is diagonal, and $A=UDV$ holds.  Let us consider an example, where $M$ is a $2\times 2$ matrix.
$$M=\left(\begin{array}{cc}
2&3\\
1&0\\
\end{array}\right)$$
Let us run the commands

\centerline{\tt SPACE = R64[]; FLOATPOS = 3; M = [[2, 3], [1, 0]];  $\backslash$SVD(M);}

The output is equal to
$$\left[
\left(\begin{array}{cc}
-0.987 & -0.16\\
-0.16 & 0.987\\
\end{array}\right),
\left(\begin{array}{cc}
3.65 & 0\\
0 & 0.822\\
\end{array}\right),
\left(\begin{array}{cc}
-0.585 & -0.811\\
0.811 & -0.585\\
\end{array}\right)\right]$$

\subsection{The Cholesky decomposition}

In general, the Cholesky decomposition is a decomposition of a Hermitian positive-definite matrix into the product of a lower-triangular matrix and its conjugate transpose, which is useful for efficient numerical solutions. It was discovered by André-Louis Cholesky for real symmetric matrices~\cite{G11}. And we also suppose that matrices are real. So, every real symmetric positive-definite matrix is equal to the product $LL^{T}$, where $L$ is a lower-triangular matrix.

The Cholesky decomposition can be calculated for a symmetric and positive definite matrix $A$ by means of the command $\mathbf{cholesky}(A)$. The result consists of two lower triangular matrices $L$ and $S$ such that $A=LL^{T}$ and $SL=I$. Let us consider an example, where $M$ is a $2\times 2$ matrix.
$$M=\left(\begin{array}{cc}
3&2\\
2&4\\
\end{array}\right)$$
Let us run the commands

\centerline{\tt SPACE = R64[]; FLOATPOS = 2; M = [[3, 2], [2, 4]];  $\backslash$cholesky(M);}

The output is equal to
$$\left[
\left(\begin{array}{cc}
1.73 & 0\\
1.15 & 1.63\\
\end{array}\right),
\left(\begin{array}{cc}
0.58 & 0\\
-0.41 & 0.61\\
\end{array}\right)
\right]$$
For large dense matrices, whose size is greater than or equal to $128\times 128$, one can use a fast algorithm $\mathbf{cholesky}(A, 1)$ that uses multiplication of blocks by the Winograd--Strassen algorithm.

\section{MODULAR ARITHMETIC}

The current version of the MathPartner service supports operations over a finite field $\mathbb{Z}/p\mathbb{Z}$, where $p$ is a prime number. One should use either {\tt SPACE = Zp[]} or {\tt SPACE = Zp32[]}. The prime number $p$ is equal to the constant {\tt MOD} or {\tt MOD32}, respectively. In the second case, $p$ satisfies the inequality $p<2^{31}$. The default value is $268435399$. For example, working over the field $\mathbb{Z}/5\mathbb{Z}$ one can run 

\centerline{\tt SPACE = Zp32[x]; MOD32 = 5; $\backslash$GCD(x+2,x-3);} 
The output is equal to $x-3$ because $-3\equiv 2\pmod{5}$. On the other hand, the same example over $\mathbb{Z}/7\mathbb{Z}$ leads to another answer.

\centerline{\tt SPACE = Zp32[x]; MOD32 = 7; $\backslash$GCD(x+2,x-3);} 
The output is equal to $1$. 

All functions using only rational operations on input data can be calculated over finite fields. In particular, for two polynomials over $\mathbb{Z}/p\mathbb{Z}$ one can calculate the greatest common divisor $\mathbf{GCD}()$ as well as Sylvester matrix $\mathbf{sylvester}()$. One can calculate the Gröbner basis of an ideal in a polynomial ring using either $\mathbf{groebner}()$ or $\mathbf{groebnerB}()$. One can also calculate the determinant $\mathbf{det}()$, echelon form $\mathbf{toEchelonForm}()$, characteristic polynomial $\mathbf{charPolynom}()$, Bruhat decomposition $\mathbf{BruhatDecomposition}()$, LDU decomposition $\mathbf{LDU}()$, and $\mathbf{LDUWMdet}()$ of a matrix.

\section{CONCLUSION}

Now the MathPartner service has become even better and allows us to solve new problems in geometry and physics. In particular, new functions allow to calculate the period of a simple pendulum as well as the circumference of an ellipse in terms of the arithmetic-geometric and the modified arithmetic-geometric means. The resultant of two univariate polynomials is a basic tool of computer algebra because it allows solving systems of polynomial equations. Matrix functions are also widely used to solve applied problems.

The reader is recommended to calculate examples of the considered quantities using the MathPartner service. These exercises will help you remember and understand the computer algebra methods better. On the other hand, new algorithms can be implemented by the user through the branch and loop operators. Moreover, the MathPartner service opens up the possibility of distance learning.



\end{document}